\documentclass[aps,prb,twocolumn,showpacs]{revtex4}
\usepackage{graphicx}
\usepackage{bm}
\usepackage{amsmath}
\usepackage[centerlast]{subfigure}
\newif\iffigs
\figstrue   
\iffigs
\fi
\def\drawing #1 #2 #3 {
\begin{center}
\setlength{\unitlength}{1mm}
\begin{picture}(#1,#2)(0,0)
\put(0,0){\framebox(#1,#2){#3}}
\end{picture}
\end{center}}

\newcommand{\St}{S}

\begin{document}
\title{Clustering and collisions of heavy particles in random smooth flows}
\author{J. Bec$^{1,2}$, A. Celani$^{2,3}$, M. Cencini$^{4,5}$ and S.
  Musacchio$^{1,4}$ } \affiliation{$^1$ Dipartimento di Fisica, Universit\`a
  di Roma ``La Sapienza'', p.zzle Aldo Moro, 2 I-00185 Roma, Italy\\ $^2$
  D\'epartement Cassiop\'ee, Observatoire de la C\^{o}te d'Azur, BP4229, 06304
  Nice Cedex 4, France.\\ $^3$ CNRS, INLN, 1361 Route des Lucioles, 06560
  Valbonne, France.\\ $^4$ SMC-INFM Dipartimento di Fisica, Universit\`a di
  Roma ``La Sapienza'' Roma ``La Sapienza'', p.zzle Aldo Moro, 2 I-00185 Roma,
  Italy\\$^5$ Istituto dei Sistemi Complessi ISC-CNR, Via dei Taurini, 19
  I-00185 Roma, Italy} \date{\today}

\begin{abstract}
  Finite-size impurities suspended in incompressible flows distribute
  inhomogeneously, leading to a drastic enhancement of collisions.  A
  description of the dynamics in the full position-velocity phase space is
  essential to understand the underlying mechanisms, especially for
  polydisperse suspensions.  These issues are here studied for particles much
  heavier than the fluid by means of a Lagrangian approach. It is shown that
  inertia enhances collision rates through two effects: correlation among
  particle positions induced by the carrier flow and uncorrelation between
  velocities due to their finite size.  A phenomenological model yields an
  estimate of collision rates for particle pairs with different sizes.  This
  approach is supported by numerical simulations in random flows.
\end{abstract}


\maketitle 
\section{Introduction}
\label{sec:intro}

The dynamics of small impurities, such as droplets, dust or bubbles,
transported by an incompressible flow is much more complex than that
of point-like fluid tracers. This is due to their finite size and to
their mass density being different from that of the carrier fluid. As
a consequence of their inertia the dynamics of such particles is
dissipative, leading to inhomogeneities in their spatial distribution.
This phenomenon, frequently referred to as {\it preferential
concentration}, has been observed for a long time in experiments (see
Ref.~\onlinecite{EF94} for a review).  Suspended particles typically
interact through collisions and chemical or biological processes. It
is therefore very important in a large spectrum of applications to
quantify the effects of inertia on these interactions.  Le us mention
for instance the problem of estimating the time scales of rain
initiation in warm clouds.\cite{PK97,FFS02,Shaw03} Other examples are
the problem of microorganism predator-prey encounters in turbulent
flows,\cite{RO88,SF00,MOPT02} and the enhancement of chemical reaction
rates for active particles suspended in fluid flows.\cite{MLG03,NTG01}
Being a common characteristics of many of the above examples, here we
focus on very dilute suspensions of very small particles that are much
heavier than the carrier fluid. They are moved by the fluid through a
viscous drag, whose nondimensional characteristic time, the Stokes
number $\St$, is proportional to the square of their radius.  In most
cases, particle sizes are below the smallest characteristic scale of
the flow, where the fluid velocity is spatially smooth.  We are
interested in the interactions taking place at those scales and we
therefore consider random smooth flows.

In the last few years, much effort has been devoted both from a
theoretical\cite{ST56,A75,WC83,KK96,ZSA03} and
numerical\cite{SC97,RC00,ZWW01,SS02a} point of view to understand and
quantify the enhancement of collision rates induced by inertia.  Two
mechanisms have been identified: preferential concentration increases
the probability for two particles to be at a colliding
distance,\cite{RC00,FFS02} detachment from fluid trajectories may
enhance the relative velocity between two approaching
particles.\cite{A75,KK96} Previous works mostly treated independently
these two mechanisms. This is justified in the
asymptotics of very small and very large inertia.  In the former,
particles are almost tracers so that their relative velocity is given
by the fluid velocity shear; the discrepancy from a uniform
distribution is responsible for an enhancement of collisions. In the
other limit particle motion is essentially ballistic: they distribute
uniformly but they may reach the same position with very different
velocities -- this is known as the {\it sling effect}.\cite{FFS02}

Several attempts have been done to bridge the gap between these two
asymptotics. For instance, Kruis and Kusters\cite{KK96} proposed an
interpolating formula for the typical particle relative velocity, by
summing together the effect of shear due the fluid velocity and the
acceleration induced by inertia. This model, though reducing to the
result of Saffman and Turner\cite{ST56} in the tracer limit and of
Abrahamson\cite{A75} for very large inertia, does not take into
account the effects of preferential concentration.  Some improvements
have been proposed in Ref.\onlinecite{ZSA03} by using two analytical
models for the fluid-particle velocities, but still neglecting the
effect of particle clustering. 

We propose here a new approach which treats these two mechanisms in a
coherent manner. Of course, we do not to provide a complete analytical
model for the collision rates, our aim is to introduce a
consistent phenomenological framework where the quantities relevant to
its estimation are stressed out. We mainly make use of tools borrowed
from dynamical systems\cite{ER85} to reinterpret inertial effects. The
basic idea is to take into account the full position-velocity
phase-space dynamics of particles. Preferential concentration of
particles is then understood as the convergence of trajectories toward
a dynamically evolving attractor in phase space. Folding of the
attractor in the velocity direction is responsible for the increase of
the velocity differences between particles.  The statistical properties of
the fractal set are determined by the carrier flow and depend on the
particle radius through the Stokes number. This approach has a natural
extension to polydisperse suspensions where particles are 
distributed over sizes: the effects of inertia on the
relative motion of particles can there be studied in terms of
correlations between attractors labeled by different Stokes numbers.

As we focus on very dilute suspensions, the dynamical effects of
collisions are here neglected, i.e.\ we work within a ghost particle
approximation.\cite{WWZ98a,WWZ98b} This allows us to investigate the
phase space in terms of the Lagrangian dynamics, that is following the
trajectories of particle pairs. Thus clustering can be characterized
in terms of the probability, $P_{1,2}(r)$, for the pair separation to
be below a distance $r$; in the same fashion we introduce the ``rate
of approach'', defined as the fraction of particle pairs at a distance
$r$ that approach each other during a unit time. This is given by the
average of the negative part of the radial velocity difference at a
separation $r$.  As we shall see, the $r$-dependence of these two
quantities sheds light on and characterizes the various regimes of
particle motion.  Moreover, the approaching rate allows to estimate,
once $r$ is fixed to be the sum of the particle radii, the collision
rates.

In the monodisperse case $P_{1,2}(r)$ behaves as a power law with an
exponent displaying a non-trivial dependence on $\St$.  This exponent
is equal to the space dimension, $d$, in both the very small and very
large Stokes number asymptotics, where particles distribute uniformly.
The scaling behavior of the rate of approach results from the joint
effect of clustering and velocity difference between particles.  The
latter behave very differently in the two asymptotics: for $\St \ll 1$
it is proportional to $r$, while for $\St \gg 1$ it becomes
independent of the separation. For intermediate values of the Stokes
number, correlations between particle separations and velocity
differences leads to non-trivial scaling behavior for the approaching
rate.

As to polydisperse suspensions we show that for the relative motion of
two particles with different Stokes numbers, a critical separation
$r_\star \propto |\St_1-\St_2|$ is singled out. Below it, the two
motions are essentially uncorrelated.  Correlations, due to the fact that
particles are suspended in the same flow, show up for length scales
above $r_\star$. The origin of this characteristic length is
understood in terms of the pair-separation dynamics, which is
dominated by the Stokes difference for $r<r_\star$ (the accelerative
mechanism), and by the fluid velocity (the shear mechanism) when
$r>r_\star$.  This crossover length separates in two distinct regimes
the scale dependence of both the probability distribution of particle
separations and their rate of approach.

Exploiting the phenomenological understanding, supplemented by 
numerical computation, of the $r$-dependence of the approaching rate (for
equal-size and different-size particle pairs) we finally propose a
semiquantitative, phenomenological model for the effective collision
kernel in polydisperse suspensions. 

The paper is organized as follows. In Section~\ref{sec:dynamics},
after recalling the equations of motion, we discuss the basic
ingredients and the validity of the approach based on Lagrangian
statistics and ghost collisions.  In Section~\ref{sec:numeric} model
flows used for numerical illustrations are described.  In
Section~\ref{sec:samestokes} and~\ref{sec:diffstokes} we investigate
the scaling behavior of the pair separation probability and of the
approaching rate in monodisperse and polydisperse suspensions,
respectively.  In Section~\ref{sec:kernel}, after a brief review of
earlier investigations and models, we propose a phenomenological model
for the effective collision kernel in polydisperse suspensions.
Section~\ref{sec:conclusions} is devoted to discussions and
conclusions.

\section{Dynamics and statistics of dilute suspensions}
\label{sec:dynamics}
The dynamics of very dilute impurities suspended in an incompressible
flow ${\bm u}$ is described by a standard model proposed by Maxey and
Riley,\cite{MR83} which was derived under the following assumptions.
The particle radius, $a$, must be much smaller than the Kolmogorov
length $\eta$. Particle Reynolds number has to be small enough to
ensure that the impurity is surrounded by a Stokes flow. The
suspension should be very dilute so that both the hydrodynamical
interactions between particles, and their feedback on the carrier
fluid can be neglected.  Since we focus on the effects induced solely
by inertia, we ignore gravity. Here we address the problem of
particles much heavier than the carrier fluid, so that this model
reduces to the Newton equation:\cite{M87}
\begin{equation}
{d^2 {\bm X} \over d t^2} + {1\over \tau} {d {\bm X} \over d t}=
{1\over \tau} {\bm u}({\bm X},t)\,,
\label{eq:motion}
\end{equation}
where ${\bm X}(t)$ denotes the trajectory.  The response time
$\tau$, frequently referred to as the Stokes time, is related to the
particle radius $a$ by
\begin{equation}
\tau= {2 \over 9}\, { \rho_p a^2 \over \rho_f \nu}\,,
\label{eq:tau}
\end{equation}
where the mass density ratio between the particle and the fluid
$\rho_p/\rho_f$ is assumed to be very large (e.g. for water droplet in
air $\rho_p/\rho_f\approx 10^3$ and for aerosols
$\rho_p/\rho_f\approx 10^3-5\times10^3$); $\nu$ is the fluid kinematic
viscosity.  It is worth reminding that, in the asymptotics of very
heavy particles, the prerequisite of small particle Reynolds number
does not restrict too much the admissible range for the values of the
Stokes time.  Indeed the nondimensional Stokes number
$\St=\tau/\tau_{\eta}=2\rho_p a^2/(9 \eta^2 \rho_f)$ (defined as the
ratio between the particle response time and the shortest
characteristic time scale of the turbulent fluid flow, i.e.\ the eddy
turnover time $\tau_{\eta}$ associated to the Kolmogorov scale) may
vary in a range from $0$, the tracer limit, to about $10$ for particle
sizes of the order of $\eta/10$ with large mass density ratio.
 
In terms of the nondimensional parameter $\St$, the equation of motion
(\ref{eq:motion}) can then be rewritten as
\begin{eqnarray}
{d {\bm X} \over d t}&=& {\bm V}\,, \nonumber\\ {d {\bm V} \over d
t}&=& {1\over \St} ({\bm u}({\bm X},t)-{\bm V})\,,
\label{eq:motion2}
\end{eqnarray}
time being now rescaled by $\tau_{\eta}$.  Here, we explicitly
introduce the particle velocity ${\bm V}$ to emphasize that, at
variance with tracers, particle dynamics takes place in the $(2\times
d)$-dimensional position-velocity phase space $({\bm x},{\bm v})$.  It
should be noted that the dynamics defined by Eq.~(\ref{eq:motion2}) is
uniformly contracting at a rate $-d/\St$, so that particle
trajectories generally concentrate in phase space onto a dynamically
evolving attractor.  In the large time asymptotics the phase-space
density of particles, solution of the Liouville equation associated to
(\ref{eq:motion2}), becomes singular with its support on this
attractor. Its statistical properties are usually
multifractal.\cite{ER85}

We consider a collection of particles with different Stokes numbers
embedded in a flow defined in a finite domain of size $L$.  Even
though the particles are carried by the same fluid flow, they converge
to different attractors according to their Stokes number.  A suitable
characterization of suspension dynamics is given by the instantaneous
phase-space densities $f_{\St}({\bm x},{\bm v},t)$. Here, $f_{\St}$ is
normalized by the total number of particles with Stokes number $\St$,
so that it can be interpreted as the probability density to find at
time $t$ a particle with Stokes number $\St$ at position ${\bm x}$ and
with velocity ${\bm v}$ and for a given fluid flow realization.

In experiments,\cite{EF94} one often has access to the distribution of
particles in the position space only.  The latter is obtained by
integrating over particle velocities, $n_\St({\bm x},t)=\int d{\bm v}
f_{\St}({\bm x},{\bm v},t)$.  For tracers, the density $n_0({\bm
x},t)$ is uniformly distributed over the domain, so that the
coarse-grained density $\tilde{n}_0(r,t)$, obtained by integrating
over small volumes of size $r$, scales as $\tilde{n}_0(r,t)\sim r^d$,
being $d$ the spatial dimension. On the contrary, for inertial
particles, typically $\tilde{n}_{\St}(r,t)\sim r^{D_1(\St)}$ with
$D_1$ markedly smaller than the space dimension $d$.  The
\textit{information dimension} $D_1$ is one among the dimensions $D_q$
that characterize the scaling properties of multifractal
densities.\cite{G83,HP83} For particle pairs dynamics the relevant
quantity is the \textit{correlation dimension} $D_2 (\St)$, which
measures the scale dependency of the probability $P_{\St}(r)$ that two
particles on the same attractor are separated by a distance smaller
than $r$. Deviations of $D_2(\St)$ from the value $d$ corresponding to
a uniform distribution are important to quantify the weight of
clustering (induced by inertia) on particle-to-particle interactions.

The \textit{radial distribution function} $g(r)$ is frequently used in
the literature to characterize clustering of inertial
particles.\cite{RC00} This quantity can be expressed as the ratio
between the actual number of particles inside an infinitesimally thin
shell of radius $r$ centered on a given particle and the number that
would be expected if the particles were uniformly distributed. It is
easily checked that the radial distribution function behaves as
$r^{D_2(\St)-d}$. For particles uniformly distributed, $D_2(\St)=d$,
so that $g(r)\sim {\cal O}(1)$. On the contrary, when $D_2(\St)<d$,
the signature of particle clustering, $g(r)$ diverges as $r\to 0$, as
it was numerically demonstrated in Refs.~\onlinecite{RC00,ZWW01}.

It is worth mentioning here that clustering is also found in tracers
advected by compressible flows. When the compressibility is
sufficiently large, particles can collapse onto point-like attractors:
this is usually referred to as the {\it strong compressibility}
r{\'{e}}gime.\cite{GV00} In such flows, particle inertia may induce a
further enhancement of clustering, leading to a strong compressibility
r{\'{e}}gime even at lower values of the compressibility.\cite{MW04}

Suspensions generally involve local interactions among the particles: dry dust
scatters through elastic or inelastic collisions depending on their impact
velocities\cite{WBC01} or water droplets coalesce to form rain
drops.\cite{FFS02} In dilute suspensions the dominant interactions are the
binary collisions taking place when two particles with radii $a_1$ and $a_2$
({\it viz.}\ Stokes numbers $\St_1$ and $\St_2$), are at a distance $r \approx
a_1+a_2$.  It has been proposed\cite{WWZ98a,WWZ98b} that for very dilute
suspensions, as considered here, the collision rates can be estimated by using
the so-called {\it ghost collisions} approach. The main idea of this method is
to let the particles overlap after they come across a distance equal to the
sum of their radii. Neither the particle sizes, nor their velocities are
modified after such an event: the collisions are not physically performed but
just recorded and counted to estimate the rate at which they happen. To
understand the relevance of such an approach let us perform the following
\textit{gedanken} experiment: we throw independently a very large number of
particle pairs inside a given domain and let them evolve in different random
realizations of the carrier flow until the two particles forming these pairs
collide. We define here the collision rate as the rate at which the number of
particle pairs which have not collided decreases with time. The dynamics
starts with transients during which the probability that two particles collide
is strongly influenced, for instance, by the choice of the initial velocities
and positions of the particles. When the dynamics of the particles has
converged to a  statistical steady state, the collision rate attains
an asymptotic value which  depends on the Stokes numbers of the
particles and on the properties of the carrier flow. This limiting
value is that measured when assuming ghost collisions. The latter are
hence expected to give a good estimate of collision rates in those
situations where the time of convergence to a steady r{\'{e}}gime is
much shorter than the mean free time between collisions of particles,
as it is for instance the case when considering very diluted
suspensions.  The framework of ghost collisions have already been
compared to several other collision schemes.\cite{WWZ98a,WWZ98b,RC00b}
As one could easily expect, it was found to be of particular relevance
when the volume fraction of particles is very low. Note finally that
beside justifying the use of this approach, the high-dilution
hypothesis has another important aspect: it permits to neglect the
feedback effect of particles on the fluid flow (the so-called reverse
coupling) and to focus on understanding the enhancement of collision
rates solely linked to particle inertia.

In order to evaluate collision rates, we introduce the rate at
which particles at a distance $r$ approach each other, which can be
written in terms of the phase-space densities as
\begin{equation}
\strut \!\!\!\!\kappa_{1,2}(r,t)\!=\left\langle \int_{|\bm x_1-\bm
x_2|=r}
\!\!\!\!\!\!\!\!\!\!\!\!\!\!\!\!\!\!\!d{\bm\Omega}_1d{\bm\Omega}_2
\,\, f_{\St_1}(\bm\Omega_1,t)\, f_{\St_2}(\bm\Omega_2,t)\,w_{1,2}^{-}
\right\rangle \,,
\label{eq:rate}
\end{equation}
where $\bm\Omega =(\bm x,\bm v)$ and the phase-space integral is
performed both on velocities and on positions satisfying $|\bm x_1-\bm
x_2|=|\bm r|=r$, and $w_{1,2}^{-}$ denotes the negative part of the
radial velocity difference $({\bm v}_2-{\bm v}_1)\cdot\hat{\bm
r}$. The average $\langle [\,\cdots] \rangle$ is performed over the
realizations of the fluid velocity field.  It easily understood that, for
$r=a_1+a_2$, the rate of approach $\kappa_{1,2}$ gives the fraction of
particles of size $a_1$ and $a_2$ colliding per unit time inside the
domain.  However, as we shall see, the full $r$-dependence of
$\kappa_{1,2}$ is very informative on the dynamics and thus
interesting by itself.

The role of inertia in increasing the rate (\ref{eq:rate}) with
respect to simple tracers can be understood easily for equal-Stokes
particles, $\St_1=\St_2=\St$. First, the clustering due to inertia
enhances the probability for particles to be close to each other. As
emphasized in Ref.~\onlinecite{FFS02}, this mechanism is very
important in the asymptotics $\St\ll 1$. Second, an additional
increase of the rate of approach $\kappa_{1,2}$ may be induced by
large radial velocity differences. As a consequence, non-trivial
dependencies of $\kappa_{1,2}$ on the scale separation $r$ will
emerge. This effect is known to be very important at large
$\St$, where caustics appear in the phase-space distribution of
particles.  Such singularities correspond to physical-space positions
where a folding of the attractor leads to the presence of different
branches in velocity at the same position, and hence, to large
velocity differences between close particles. Such a folding in
velocity is the basic mechanism leading to what is known as the sling
effect.\cite{FFS02}

\subsection*{Lagrangian statistics} 
\label{sec:lagrangian}
Clearly, a direct $(2\times d)$-dimensional numerical integration to
determine phase-space densities is a difficult task. We thus adopt a
different strategy by using a Lagrangian approach.  Indeed,
investigating two-point correlations of densities, as needed for
binary collisions, is equivalent to study the relative motion of
particle pairs.  Moreover the average over velocity configurations can
now be replaced by a time average.\cite{nota}.  To
evaluate the rate of approach $\kappa_{1,2}$, it is sufficient to
follow two particles with Stokes numbers $\St_1$ and $\St_2$, and to
compute the time average of the negative component of the radial
velocity difference conditioned on the pair distance,
$r$. Numerically, it is more convenient to condition such averages on
having the two particles at a distance smaller than $r$, rather than
exactly equal to $r$. This amounts to work with cumulative quantities
that are by far less noisy. We hence define
\begin{eqnarray}
{\cal K}_{1,2}(r) =\overline{
|{W}_{1,2}| \!\!\!\!\phantom{ 1^{1^1}} \!\!H(- {W}_{1,2})\,H(r-{ |{\bm
    R}_{1,2}|})}
\,,
\end{eqnarray}
where ${\bm R_{1,2}}={\bm X}_1-{\bm X}_2$, ${W_{1,2}}=({\bm V}_1-{\bm
  V}_2)\cdot{\bm R_{1,2}}/|{\bm R_{1,2}}|$, $H$ is the Heaviside function and
the overbar denotes Lagrangian time averaging; the rate (\ref{eq:rate}) is
then obtained as $\kappa_{1,2}(r)=\partial_r {\cal K}_{1,2}(r)$.

To quantify the balance between the two effects enhancing collisions
mentioned above, namely clustering and large velocity differences, we
make use of the probability density function $p_{1,2}(r)$ for the pair
separation. It is clear that this quantity is related to the radial
distribution function by $g_{1,2}(r) = L^d p_{1,2}(r)/r^{d-1}$. Similarly,
instead of measuring directly $p_{1,2}(r)$, we consider the cumulative
probability
\begin{equation}
P_{1,2}(r)=\overline{H(r-{ |{\bm R}_{1,2}\!\!\!\phantom{1^{1^1}}
    \!\!\!\!|})}\,,
\end{equation}
that is the probability that two particles are at a distance closer
than $r$. It is clear that $p_{1,2}(r)=\partial_r P_{1,2}(r)$ (note that
while $P_{1,2}$ is adimensional, $p_{1,2}$ has the dimension of the 
inverse of a length.) 

For separations $r$ much smaller than the smallest
characteristic length scale of the fluid flow (e.g.\ the Kolmogorov
scale $\eta$ for turbulent flows), both
$p_{1,2}$ and $\kappa_{1,2}$ are expected to display power law
behaviors:
\begin{eqnarray}
p_{1,2}(r) \sim \left({r \over L}\right)^{\mu(\St_1,\St_2)}\,, \\
\kappa_{1,2}(r)\sim \left({r\over L} \right) ^{\gamma(\St_1,\St_2)}\,.
\end{eqnarray}
Therefore, the corresponding cumulative quantities $P_{1,2}$ and ${\cal
K}_{1,2}$ do also behave algebraically with exponents $\mu+1$ and
$\gamma+1$. Note that these quantities are symmetric under particle
exchange $(1,2)\mapsto(2,1)$, so that $\mu(\St_1,\St_2) =
\mu(\St_2,\St_1)$ and $\gamma(\St_1,\St_2) = \gamma(\St_2,\St_1)$.

For equal-Stokes particles ($\St_1\!=\!\St_2\!=\!\St$),
$\mu(\St)\!+\!1$ is exactly the correlation dimension $D_2(\St)$ of
the projection of the attractor onto the physical space. For generic
pairs, the exponent $\mu(\St_1,\St_2)+1$ measures the correlation
between the distributions of particles of the two types.  For smooth
flows, the exponent $\gamma$ is constrained between $\mu$ and $\mu
+1$. The lower bound is attained when particle velocities and their
separation $r$ are uncorrelated, while the upper bound correspond to a
velocity difference between particles proportional $r$.

\section{Model flows}
\label{sec:numeric}
Since collisions occur at distances comparable with the particle
radii, therefore typically much smaller than the turbulent viscous
scale $\eta$, we can limit ourselves to smooth incompressible fluid
velocity fields.  In order to perform long time averages at a
reasonable cost, we consider synthetic random flows.  The physical
mechanisms are not expected to be very sensitive to the space
dimensionality. The only important effect is that the probability of
finding two close particles decreases with the dimension $d$. 
We therefore choose two-dimensional flows in a finite domain of size $L$. 
The validity of the results presented in this work should extend, at least
qualitatively, to more realistic velocity fields.

First, we consider an isotropic, homogeneous and Gaussian flow.  The Fourier
modes of the fluid velocity are Ornstein--Uhlenbeck processes satisfying
\begin{equation}
{d\hat{\bm u}_{\bm k} \over dt} = -\frac{1}{\tau_f} \hat{\bm u}_{\bm
  k} +c_{\bm k} \bm \xi_{\bm k}\,,
\label{eq:ou}
\end{equation}
where the ${\bm \xi}_{\bm k}$'s are independent white noises and the
correlation time $\tau_f$ is independent of the wavevector ${\bm k}$.
The process (\ref{eq:ou}) is defined over $8$ modes and the constants
$c_{\bm k}$'s have been suitably chosen to ensure statistical isotropy
at small scales.  This velocity field, being smooth in space and
continuous in time, is expected to mimic the dissipative-range
dynamics, when neglecting intermittency effects; $L$ should then be
understood as the Kolmogorov scale $\eta$, and the correlation time
$\tau_f$ as the Kolmogorov time $(\eta^2/\epsilon)^{1/3}$.  Note that
the existence of random dynamical attractors for heavy particles, and
hence of a statistically steady state, has been recently proved for
this model flow.\cite{SS02b} The advantage of using a reduced number
of modes is that the Fourier summation can be directly performed at
particle positions, enabling to resolve the finest scales of the
dynamics without interpolation.  Particles evolve according to
Eq.~(\ref{eq:motion}) in a periodic domain of size $L \times L$ (here
$L=2\pi$).  Time marching is performed by a fourth-order Runge-Kutta
scheme for times up to $10^6-10^7 \, \tau_f$.  The Stokes number is
defined as $\St=\tau/\tau_f$.

Second, we investigate the alternating shear flow:
\begin{eqnarray}
u_1\!\! &=&\!\! 0 \;;\; u_2={U_n \, x_1(L-x_1) \over L^2} \quad {\mbox
{for}} \;\; t\in [{\scriptstyle (n-1)}T,{\scriptstyle (n-{1\over
2})}T]\,,\nonumber\\ u_1\!\! &=& \!\!{U_n \, x_2(L-x_2) \over L^2}
\;;\; u_2=0 \quad {\mbox {for}}\;\; t\in [{\scriptstyle (n-{1\over
2})}T,{\scriptstyle n}T]\,,
\label{eq:shear}
\end{eqnarray}
where periodic (of period $L$) boundary conditions are imposed in
$(x_1,x_2)$, and $U_n=\pm U$ is chosen randomly at each time interval
$T$ with equal probability.  In this case the Stokes number is defined
as $\St=\tau/T$. The reason for studying such a model flow is
twofold. First, it is interesting for testing the robustness of the
results, at least at a qualitative level. Second, this model flow has
the nice feature that the integration of particle motion can be done
explicitly in each time interval of duration $T/2$, allowing for
extremely long time averages (up to $10^8-10^{9}\, T$).

\section{Local dynamics of monodisperse suspensions}
\label{sec:samestokes}

We start considering the asymptotics of small Stokes numbers, that is
when the particle response time is much smaller than any typical time
scale of the flow. A simple perturbative expansion of (\ref{eq:motion2})
implies that inertial particles behave as fluid tracers evolving in a
slightly compressible flow whose divergence is proportional to
$\St$.\cite{M87} At small separations $r\ll L$, being the carrier flow
spatially smooth at these scales, the radial velocity difference of
two particles is proportional to $r$. Hence, for equal-size particles
$\gamma=\mu+1$. For a vanishing Stokes number $\St = 0$, the particles
behave as tracers in an incompressible flow, therefore they distribute
uniformly in space, meaning that $\mu=d-1$ and $\gamma=d$.  For small
yet finite $\St$, if the velocity field is isotropic and homogeneous,
one expects the discrepancy from the uniform distribution to behave as
$\St^2$, and in particular $\mu(\St)\simeq d-1-\alpha \St^2$ ($\alpha$
is a flow-dependent constant).\cite{BFF01,ZA03} In terms of the radial
distribution function this means that $g(r)$ diverges in the limit
$r\to 0$ as $r^{-\alpha \St^2}$.  This behavior has been indeed
observed in Gaussian, random incompressible flows\cite{B03} and in
direct numerical simulations of three-dimensional turbulent
flows.\cite{FP04} Therefore, in the case of small Stokes numbers, the
main effects of inertia on particle interactions stems from
preferential concentration, which increases the probability of finding
close particles.\cite{FFS02,FP04}

\begin{figure}[b!]
  \iffigs \includegraphics[scale=.4]{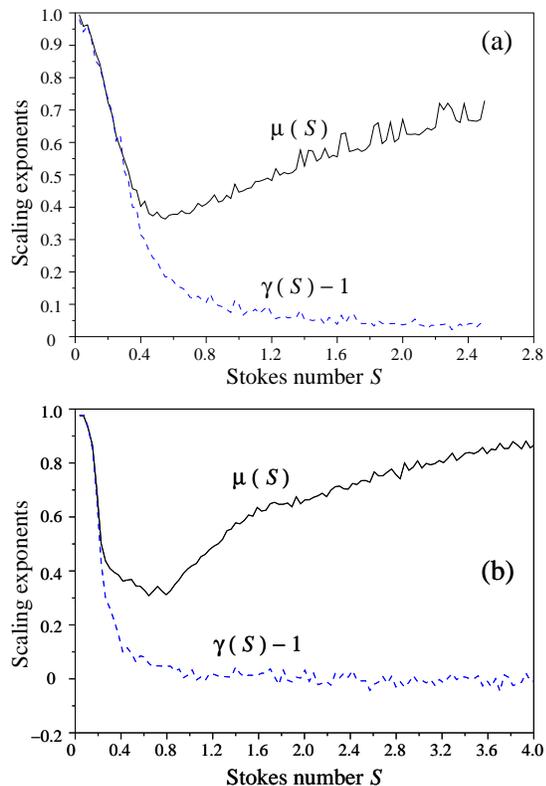}\\
  \else 
  {\drawing 60 10 {Fig1a} }
  {\drawing 60 10 {Fig1b} }\fi
  \caption{\label{fig:d2_gamma} (a) Scaling exponents $\mu$ of the density
    correlations and $\gamma$ of the rates of approach as a function of the
    Stokes number for the time-correlated Gaussian random flow. Time averages
    are performed over $5\times10^6\,\tau_f$ (b) Same as in (a) for the shear
    flow with $T\!=\!L=\!1\,,\,U\!\!=\!\!\sqrt{23}$. Time averages are over
    $2\!\times\!\!10^9\, T$.  In both cases the exponents are measured as the
    mean logarithmic derivative of ${\cal K}$ over two to three decades in
    $r$. }
\end{figure}

The other asymptotics is also easy to understand: for $\St \gg 1$,
particles have a nearly ballistic motion. They are thus expected to
distribute uniformly in the whole domain, i.e. $\mu \sim d-1$.
Moreover, even close particles have typical velocity differences that
do not depend on their separation, i.e.\ $\gamma\sim \mu \sim
d-1$. Therefore, even though in both large and small Stokes numbers
asymptotics particle densities are uniform, the rates of approach are
markedly different.

The exponents $\mu$ and $\gamma$ computed numerically for the two
flows described in Section~\ref{sec:numeric} are shown in
Fig.~\ref{fig:d2_gamma} as a function of the Stokes number $\St$.  At
both small and large values of $\St$ the expected behaviors are rather
clear: when $\St\ll 1$, there is coincidence between $\gamma-1$ and
$\mu$ (velocity differences are proportional to the separation), while
for $\St\gg1$, one observes that $\gamma \to 2$ (velocities are almost
uncorrelated). For intermediate Stokes numbers non-trivial
dependencies between particle separations and velocity differences
appear. In particular, notice that $\mu$ has a minimum (meaning
maximum of clustering) for a finite value of $\St$ that, as far as we
know, cannot be predicted by any present theory.  Moreover, close to
this minimum $\gamma-1$ noticeably deviates from $\mu$. As we shall
see in Sec.~\ref{sec:kernel}, this implies a further increase of the
collision rate in this range of Stokes.  It is also worth stressing
that both models lead to qualitatively similar results, bringing
evidence for the robustness of these features. Note that this property
Note that this property has already been observed numerically and
experimentally.\cite{EF94}
\begin{figure}[hb!]
  \iffigs \includegraphics[scale=.6]{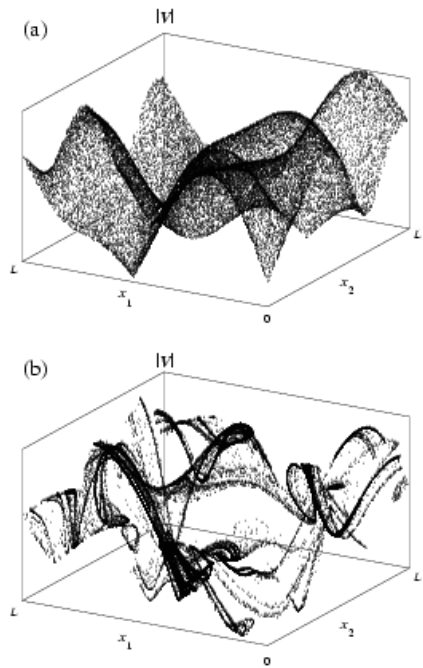}\\
  \else 
  {\drawing 60 10 {Fig2a} }
  {\drawing 60 10 {Fig2b} }\fi
  \caption{\label{fig:3bis} Modulus of the particle velocities as a
    function of their positions for two different values of the Stokes
    number (a) $\St =10^{-3}$ and (b) $\St=1$. At small $\St$ the
    surface identified by the particles velocities is very close to
    that of the modulus of the Eulerian velocity field, meaning that
    at fixed spatial position the distribution of the particle
    velocity is sharply peaked on the fluid one. For larger Stokes
    numbers, the attractor folds in the velocity direction
    allowing particles to be very close with very different
    velocities.}
\end{figure}

The non-trivial relations between $\gamma$ and $\mu$ at varying $\St$
can be understood in terms of two competing effects in phase space:
the folding of the attractor in the $\bm v$-direction and its tendency
to contract toward the surface defined by the instantaneous fluid
velocity.  The typical rate of this relaxation is given by the Stokes
time.  For $\St = 0$ it becomes infinitely fast, preventing folding so
that the particle velocity field is mostly mono-valued (see
Fig.~\ref{fig:3bis}a).  As the Stokes number increases the probability
of finding particles at the same position with different velocities
becomes larger (see Fig.~\ref{fig:3bis}b).  Such
points, once projected onto position space, lead to what can be seen
as self-intersections of the attractor projection.  This phenomenon was
pointed out as the ``sling effect'' in Ref.~\onlinecite{FFS02}.  We
stress that the numerical observations reported in
Fig.~\ref{fig:d2_gamma} show that this phenomenon starts to be
important already for relatively small Stokes numbers, as signaled by
the deviation of $\gamma-1$ from $\mu$.

To summarize, as we expected, the two quantities of interest to
measure clustering and collisions have, for particles with the same
Stokes number $\St$, a power-law behavior at small separations $r$:
\begin{eqnarray}
p_{\St}(r) &\sim& {{\rm C}_\St\over L} \left({r \over L}\right
)^{\mu(\St)}\,, \label{eq:PsameSt} \\ \kappa_{\St}(r) &\sim& {{\rm
C}_\St {\rm V}_\St \over L} \left({r \over L}\right
)^{\gamma(\St)}\,. \label{eq:KsameSt}
\end{eqnarray}
The constant ${\rm C}_\St$ depends on $\St$ and on the statistics of
the velocity gradients of the carrier fluid.  For three-dimensional
turbulent flows it may also depend on the Reynolds number.  ${\rm
V}_\St$ is a typical velocity of the particles with Stokes number
$\St$. For small $\St$ it is of the order of the root-mean-square
velocity of the carrier flow, $u_{rms}$.  For $\St\gg 1$ it can be
shown\cite{ZA03} that ${\rm V}_\St\sim u_{rms} \St^{-1/2}$.

\section{An extension to polydisperse suspensions}
\label{sec:diffstokes}
We now investigate suspensions with particles having different Stokes
numbers.  To this aim we consider the equations governing the relative
motion of two particles with Stokes numbers $\St_1$ and $\St_2$.  The
separation $\bm R$ and the relative velocity $\bm W$ evolve according
to
\begin{eqnarray}
  {d {\bm R} \over d t} &=& {\bm W}\,, \nonumber\\ {d {\bm W} \over d
  t} &=& {1 \over \tau}\, {{\Delta{\bm u}-{\bm W}}\over 1-{\theta^2 /
  4}} - \frac{\theta}{\tau} \,{\bar{\bm u}-\bar{\bm V} \over
  1-{\theta^2 / 4}}\,,
  \label{eq:relativmotion}
\end{eqnarray}
where $\bar{\bm u}=({\bm u}_1+{\bm u}_2)/2$, $\Delta{\bm u}={\bm u}_1-{\bm
  u}_2$ and $\bar{\bm V} = ({\bm V}_1 + {\bm V}_2)/ 2$.  Here and in the
sequel, $\tau=(\tau_1+\tau_2)/2$, $\St=(\St_1+\St_2)/2$ and $\theta
=(\tau_1-\tau_2)/{\tau} =\Delta\St/{\St}$.

\begin{figure}[tbp]
  \iffigs \includegraphics[scale=.5]{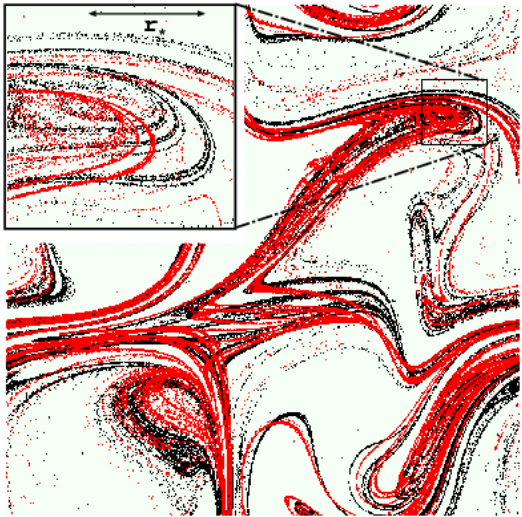}
  \else \drawing 60 10 {Fig3} \fi
  \caption{\label{fig:snapshotdiffstokes} Snapshot of the positions of
    $N=4\times 10^5$ particles associated to two different values of the
    Stokes number, $\St=0.4$ (black) and $\St=0.5$ (gray, red online) for the
    random flow (\ref{eq:ou}). The upper-left inset shows a zoom illustrating
    the effects induced by the difference in Stokes number (see text for
    details). }
\end{figure}

The two terms in the right-hand side of (\ref{eq:relativmotion}) are
associated to different effects.  The first corresponds to the relaxation of
the relative velocity to the fluid velocity difference. The second is
proportional to the difference in Stokes numbers, and therefore it vanishes
for equal-size particles. The characteristic length-scale 
\begin{equation}
r_\star = L |\theta|\,,
\label{def:rstar}
\end{equation}
distinguishes two ranges of scales. When $R = |\bm R| > r_\star$, the
particle velocity difference is driven by the fluid velocity difference
$|\Delta {\bm u}| \sim R$ while when $R < r_\star$ it is driven by the term
originating from the difference in particle response time. As a
consequence for $R < r_\star$ particle motion is uncorrelated, while
above $r_\star$ correlations induced by the fact that particles are
transported by the same velocity field show up. This mechanism is
exemplified in Fig.~\ref{fig:snapshotdiffstokes}, where the
simultaneous snapshots of two populations of particles characterized
by different Stokes numbers are shown: at large scales the two
distributions look essentially the same, while a closer inspection
reveals their differences below the crossover length.

Two cases need to be distinguished: small Stokes number differences,
corresponding to $r_\star \ll L$ for which an intermediate asymptotic range is
present, and finite Stokes number differences for which $r_\star \sim {\cal
O}(L)$, so that the intermediate range is absent.

\bigskip
\noindent\textit{a.\ Small Stokes number differences} ($\theta\ll
1$)\par\medskip
In the intermediate asymptotic range $r_\star \ll r \ll L$, the
particle separation is essentially governed by the first term in
(\ref{eq:relativmotion}). The quantities of interest are thus
described by the two-point dynamics associated to a single Stokes time
-- the mean Stokes time. The geometrical interpretation of this
dynamical effect is evident from the snapshot of particle positions
shown in Fig.~\ref{fig:snapshotdiffstokes}.  Indeed at scales larger
than $r_\star$ the effects of polydispersion are negligible. In this
range of scales we recover the results from the previous Section,
namely $p_{1,2}(r) \approx p_{{\St}}(r)$ and $\kappa_{1,2}(r) \approx
\kappa_{{\St}}(r)$ given by (\ref{eq:PsameSt}) and (\ref{eq:KsameSt}),
respectively.

As displayed in the inset of Fig.~\ref{fig:snapshotdiffstokes} the
situation is quite different for $r \ll r_\star \ll L$. Now, the time
derivative of $\bm W$ does not depend on $\bm R$, but only on the
differential acceleration induced by distinct Stokes numbers.  Because
of the resulting relative shift between the two attractors in phase
space, the velocity difference is independent of the separation. As a
consequence, impurities with different $\St$ see each other, at such
scales, as a gas of uniformly distributed free-streaming particles.
This leads to
\begin{eqnarray}
  p_{1,2}(r) &\sim& {{\rm C}_{{\St}}  \left|\theta
  \right|^{\mu(\St)-d+1}\over L} \left({r \over L}\right )^{d-1}\,,
  \label{eq:PdiffSt} \\ \kappa_{1,2}(r) &\sim& {{\rm C}_{{\St}}\, 
{\rm V}_{{\St}}
    \left|\theta \right|^{\gamma(\St)-d+1} \over
    L} \left({r \over L}\right)^{d-1}\,.
  \label{eq:KdiffSt}
\end{eqnarray}
These expressions match the intermediate asymptotics
(\ref{eq:PsameSt}) and (\ref{eq:KsameSt}) for $r=r_\star$. The
constants ${\rm C}_{{\St}}$ and ${\rm V}_{{\St}}$ are related to the
two-point motion associated to the mean Stokes number
${\St}=(\St_1+\St_2)/2$.

\begin{figure}[t!]
\iffigs \includegraphics[draft=false, scale=.58, clip=true]{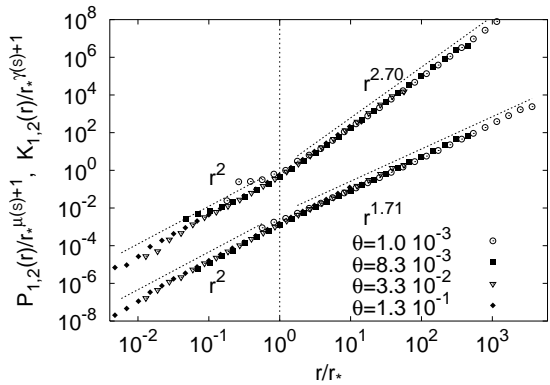}
  \else \drawing 60 10 {Fig4} \fi
\caption{Rescaled cumulative probability $P_{1,2}(r)$ that two particles
  are at a distance smaller than $r$ (lower curve), and cumulative approaching
  rate ${\cal K}_{1,2}(r)$ (upper curve) as a function of $r/r_\star$.
  Results are shown for $\St=0.18$ for different values of $\theta$ as in the
  label.  Note the collapse of the different curves; the dashed lines indicate
  the scaling behavior below and above $r_\star$. It is worth also noticing
  that for $r<r_{\star}$ both $P$ and ${\cal K}$ scale as $r^2$ while for
  $r>r_{\star}$ they scale as $r^{\mu(\St)+1}$ and $r^{\gamma(\St)+1}$,
  respectively. As one can see for such a small values of $\St$ $\gamma
  \approx \mu+1$. Data refer to simulations in the random shear flow
  (\ref{eq:shear}) with $T=1\,,\;L=1\,,\;U=\sqrt{23}$. The curve for $P_{1,2}$
  has been shifted down for plotting purposes.}
\label{fig:shear_rstar}
\end{figure}

The presence of a characteristic length scale $r_\star$ separating
these two regimes is confirmed by numerical experiments  
as shown in Fig.~\ref{fig:shear_rstar}.

\bigskip
\noindent\textit{b.\ Large Stokes number differences} ($\theta =
{\cal O}(1)$) \par\medskip

In this case $r_\star$ is of the order of the domain size $L$ and the
intermediate asymptotic regime disappears.  Therefore, at any scale $r \ll L$
each of the two particles sees the other as if uniformly distributed with
independent velocity.  The probability that two particles are at a distance
$r$ may thus be written
\begin{equation}
  p_{1,2}(r) \sim {{\rm C}_{\St_1,\St_2} \over L}\left({r \over
  L}\right )^{d-1}\,.
  \label{eq:PverydiffSt}
\end{equation}

Regarding the rate of approach, the velocity difference can
be approximated as the difference between two uncorrelated
velocities. This yields to estimate its typical value as $({\rm
  V}_{\St_1}^2+{\rm V}_{\St_2}^2)^{1/2}$, where ${\rm V}_{\St_j}$ is
the characteristic velocity of particles with Stokes number $\St_{j}$.
It follows
\begin{equation}
  \kappa_{1,2}(r) \sim {{\rm C}_{\St_1,\St_2} \over L}\,\left({\rm
    V}_{\St_1}^2+{\rm V}_{\St_2}^2\right)^{1/2} \left({r \over
    L}\right)^{d-1}\,.
  \label{eq:KverydiffSt}
\end{equation}

\section{Phenomenological model for the collision kernel}
\label{sec:kernel}
Let us, for the sake of simplicity, focus on polydisperse suspensions of
particles having the same mass density. This assumption implies a one-to-one
correspondence between the Stokes number and the particle size, namely $\St
\propto a^2$. In this framework the simplest phenomenological description of
particle collisions in dilute suspensions focuses on the total number of
collisions per unit time between particles of sizes $a_1$ and $a_2$ averaged
over the fluid velocity realizations:
\begin{equation}
  {\cal N}_c(a_1,a_2) = Q(a_1,a_2)\, {\cal N}_1\,{\cal N}_2,
  \label{eq:defcoll}
\end{equation}
where ${\cal N}_{i}$ is the mean number of particles with size $a_{i}$
inside the domain.  The effective collision kernel, $Q(a_1,a_2)$, is
given by the average rate at which particles associated to Stokes
number $\St_1\propto a_1^2$ and $\St_2\propto a_2^2$ arrive at a
distance equal to the sum of their radii, i.e.\ $Q(a_1,a_2)=
\kappa_{1,2}(a_1+a_2)$.  Building quantitative models of the collision
kernel in turbulent flows is of utmost importance for many natural
phenomena and industrial processes. For instance, we stress that the
knowledge of $Q$ is crucial to understand the evolution of the droplet
size distribution in clouds,\cite{PK97} which is of paramount
importance for a theoretical comprehension of the rain drops
formation.\cite{FFS02} We believe that the main ingredients needed for
such a program are, at least at qualitative level, independent on the
complexity of the flow. Therefore, here we shall concentrate on the
simpler case of random flows, ignoring some (though important) aspects
of turbulent flows. Our aim is to develop a semi-quantitative
phenomenological model for $Q(a_1,a_2)$ able to capture the main
ingredients of collisions in suspensions with a broad distribution of
particle sizes.

There are two asymptotic regimes where the properties of $Q$ are well
understood, namely for vanishing inertia ($\St\to 0$) and for infinite
response time ($\St\to \infty$).  As stated in the previous Sections,
in both limits particles distribute uniformly, so that investigating
the collision kernel reduces to understanding the statistics of
velocity differences. 

Following Saffman and Turner\cite{ST56}, in smooth flows and for $\St\to 0$
the collision kernel can be expressed as
\begin{equation}
  Q_{ST}(a_1,a_2) = {\rm D}\,\lambda\,\left(({a_1+a_2})/L\right)^{d}\,.
  \label{eq:saffmanturner}
\end{equation}
This is obtained by multiplying the geometrical cross section $\propto
(a_1+a_2)^{d-1}$ with the typical velocity difference between the two
particles. The latter is approximated by $\lambda (a_1+a_2)$, being $\lambda$
the characteristic fluid velocity gradient; in turbulent flows the latter 
is usually estimated as $\lambda \approx (\epsilon/\nu)^{1/2}$. 

For $\St\to \infty$, it was suggested by
Abrahamson\cite{A75} that the kernel can  be obtained by a molecular-chaos
type of argument, because positions and velocities of particles are
uncorrelated:
\begin{equation}
  Q_{A}(a_1,a_2) = \frac{{\rm D}'}{L^d}\, \left({\rm V}^2_{\St_1}+{\rm
    V}^2_{\St_2}\right)^{1/2}\, \left(a_1+a_2\right)^{d-1}\,, 
  \label{eq:abrahamson}
\end{equation}
where ${\rm V}_{\St} \sim u_{rms}/\St^{1/2}$.  

In the last few years most of theoretical efforts focused on the
intermediate regime, for which many models and predictions have been
proposed.  For example, in the regime $\St \ll 1$ of very small Stokes
numbers, collision rates are enhanced solely by preferential
concentration through an increase in the probability of having
particles at a colliding distance. This was indeed demonstrated by
means of DNS and theoretical
analysis.\cite{WWZ98b,RC00,BFF01,FFS02,ZSA03} Thus most of the efforts
focused in predicting the net effect of clustering.\cite{FP04} This is
surely crucial for (close to) monodisperse suspensions of particles
with very small sizes, but this approach cannot catch collisions between
particles with different sizes.  Moreover, even for same-size
particles, as soon as the Stokes number reaches small but finite
values, inertia besides increasing the probability to find close
particles also enhances particle relative velocities.  This is clear
from Fig.~\ref{fig:d2_gamma} where we see that, as $\St$ approaches
the value where clustering is maximal, the particle velocity
difference is no more proportional to the particle separation. This
leads to an additional increase in the collision rates.  Indeed, from
comparisons with DNS, Wang {\it et al.}\cite{WWZ00} argued that most
models are not able to accurately predict the relative velocity.  As
to collisions between particles having different sizes the situation
is even more involved. Clearly as soon as the Stokes number difference
is not negligible the accumulation effect induced by inertia need to
be understood in terms of correlations among particle positions
belonging to different populations (what we called correlations
between different attractors). As far as we know, only few attempts in
this direction have been considered. Among them we mention Zhou {\it
et al.}\cite{ZWW01} who investigated by means of DNS, bidisperse
suspensions in frozen turbulent flows.  In their study they found a
reduction of the accumulation effect and an increase in relative
velocity. In particular, they proposed a simple model for the particle
velocities correlation, in which they prescribed an extremely
simplified fluid velocity correlation. Though in fairly good agreement
with simulations, their results are rather difficult to interpret
because spatial correlations of the fluid velocity are not taken into
account. Moreover, their approach seems to be justified only for
particle response time of the order of the correlation time of the
large scales, $T_e$. Indeed, they consider in their numerical
simulations particles with response
times up to $3T_e$. Also Kruis and Kusters\cite{KK96} proposed a model
for bidisperse collisions. In their approach they separate, in a
somehow artificial way (as argued in Ref.~\onlinecite{ZSA03}), two
contributions.  The first named shear mechanism corresponds to the
carrier fluid velocity shear, that is very similar to the
Saffman-Turner result.  The second, deriving from the acceleration
induced by the different inertia, was called accelerative
mechanism. Also this model appear of difficult interpretation due to
the absence of spatial correlation in its derivation.  Moreover, both
mechanisms are considered to be effective simultaneously and the
dominance of one over the other is only due to intensity of the
velocity gradients, while clearly the particle sizes should play a
role too.

In the sequel, exploiting the results obtained in the previous
sections, we propose a phenomenological model of the collision kernel
which reduces to the known results in the two asymptotics of vanishing
and very large Stokes number. In the intermediate regime, an
interpolation between the two asymptotics naturally emerges in terms
of the exponent $\gamma(\St)$ that, (as seen in
Fig.~\ref{fig:d2_gamma}) out of the two limiting regimes, cannot be
trivially related to $\mu$. Moreover, as we shall see, the
identification of the crossover scale $r_\star$ (\ref{def:rstar})
provides a natural way to discern between differential acceleration
and shear mechanisms.  To be used for quantitative predictions an
analytical estimation of $\gamma(\St)$ would be needed. Even for
relatively simple random flows, this is far from the present
capabilities. However, this quantity is accessible in simulations and
may be in principle measured from experiments where particle tracking
can be done with a high accuracy.

\begin{figure}[t!]
\iffigs \includegraphics[scale=.62]{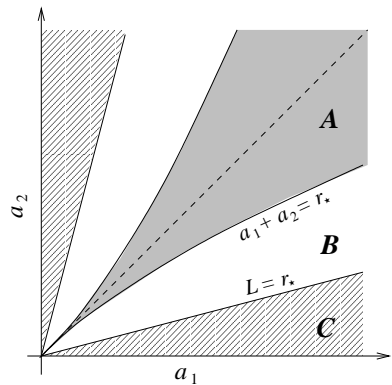}
  \else \drawing 60 10 {Fig5} \fi
\caption{\label{fig:sketchkernelzones} Sketch of the different regions
  in the $(a_1,a_2)$ plane corresponding to different contributions of
  $\kappa_{1,2}$ to the effective collision rates.}
\end{figure}

According to the different cases described in Sec.\
\ref{sec:diffstokes}, the $(a_1,a_2)$ plane has to be divided into
different regions corresponding to the various behaviors of
$\kappa_{1,2}(r)$ evaluated at $r = a_1 +a_2$.  Three regions can be
identified (see Fig.~\ref{fig:sketchkernelzones} for a sketch): region
$\bm A$ where $r_\star \le a_1+a_2\le L$, $\bm B$ that is defined by
$a_1 +a_2 \le r_\star \le L $, and $\bm C$ by $a_1+a_2\le L \le
r_\star$.  Reminding that $r_\star \propto
|a_1^2-a_2^2|/(a_1^2+a_2^2)$, and the results of
Sects.~\ref{sec:samestokes} and \ref{sec:diffstokes} one has the
following behaviors for the collision rates.

\smallskip
\noindent{$\bm A$-}  In the gray region defined by
\begin{equation}
 |a_1-a_2| \le 2\,L - \left(4\,L^2 - (a_1 + a_2)^2\right)^{1/2}\,,
  \label{eq:defrange2}
\end{equation}
the behavior of the mean radial velocity difference is well approximated
by the two-point motion of particles with the mean Stokes number $\St
\propto (a_1^2+a_2^2)$. The collision rate thus reduces to that of
particles with the same Stokes number, namely from Eq.~(\ref{eq:KsameSt})
\begin{equation}
  Q(a_1,a_2) \sim {\rm C}_{\St}\,{\rm V}_{\St}\,
  \frac{(a_1+a_2)^{\gamma(\St)}} {L^{\gamma(\St)+1}}\,,
  \label{eq:raterange2}
\end{equation}
where the constant ${\rm C}_{\St}$ depends on the Stokes number and on
the fluid velocity statistics.
\smallskip

\noindent{$\bm B$-} In the white region, the
inequality~(\ref{eq:defrange2}) is fulfilled, yet it holds 
\begin{equation}
  |a_1-a_2| \le (2-\sqrt3)\, (a_1 + a_2)\,.
  \label{eq:defrange1}
\end{equation}
The motions of the two particle become
uncorrelated and from Eq.~(\ref{eq:KdiffSt}) one obtains:
\begin{equation}
\strut \!\!\!\!  Q(a_1,a_2) \sim {\rm C}_{\St}\,{\rm V}_{\St}\, \frac{
  |a_1-a_2|^{\gamma(\St)-d+1} \,(a_1+a_2)^{\gamma(\St)}}
  {(a_1^2+a_2^2)^{\gamma(\St)+1}}.
  \label{eq:rateinterrange}
\end{equation}

\smallskip
\noindent{$\bm C$-} In the hatched region where the
inequality~(\ref{eq:defrange1}) is not satisfied, the collision
rate is given by
\begin{equation}
\strut\!\!\!\!  Q(a_1,a_2) \sim {\rm C}^{'}_{\St_1,\St_2}\,\left({\rm
  V}_{\St_1}^2+{\rm V}_{\St_2}^2\right)^{1/2}\,\frac{(a_1+a_2)^{d-1}}{L^d}\,,
  \label{eq:raterange1}
\end{equation}
where ${\rm V}_{\St_1}$ and ${\rm V}_{\St_2}$ are the typical velocities
associated to particles of size $a_1$ and $a_2$, respectively. Of course,
close to the boundary of the hatched region, the constant ${\rm
  C}^{'}_{\St_1,\St_2}$ should be suitably chosen to ensure continuity of the
collision kernel $Q$ in the plane $(a_1, a_2)$.  Note that this expression is
consistent with Abrahamson prediction (\ref{eq:abrahamson}), which is
expected to hold for large values of the Stokes number.
\smallskip

We now make some comments on the different forms taken by the kernel,
accordingly to the values of particles sizes.  The first obvious
information is that Eq.~(\ref{eq:raterange2}) comprises the two
asymptotics of very small (\ref{eq:saffmanturner}) and very large
Stokes numbers (\ref{eq:abrahamson}) when $\gamma$ is replaced by its
limiting values.  An important observation is that collisions between
particles with different Stokes numbers are related to non-trivial
scaling behavior only when their radii are rather similar (i.e.\ in
the gray area of Fig.\ \ref{fig:sketchkernelzones}).  In this region
the rates can be obtained in terms of the dynamics for particles with
the mean Stokes number. The main information is contained in the
Stokes-number dependence of the exponent $\gamma$.

\begin{figure}[t!]
  \iffigs \includegraphics[scale=.395]{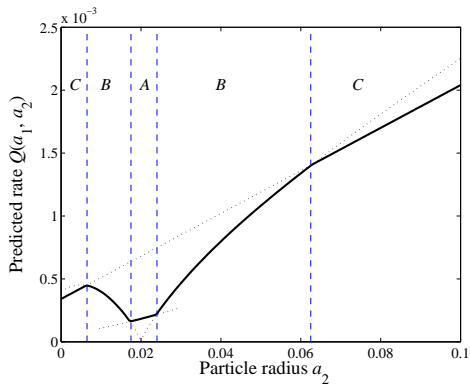}
  \else \drawing 60 10 {Fig6} \fi
  \caption{\label{fig:predicted} Typical functional shape (bold line)
    of the collision kernel, represented here as a function of $a_2$
    through a cut with $a_1=0.02$ fixed.  We chose here $\gamma=1.8$
    and the numerical factors and constants to fit the order of
    magnitude obtained in our numerical experiments.  The effective
    kernel takes one of the different functional forms
    (\ref{eq:raterange2}), (\ref{eq:rateinterrange}) or
    (\ref{eq:raterange1}) (represented as dotted lines), depending on
    whether $(a_1,a_2)$ is in the region $\bm A$, $\bm B$ or $\bm C$.}
\end{figure}
According to the predictions (\ref{eq:raterange2}),
(\ref{eq:rateinterrange}) and (\ref{eq:raterange1}) the
typical shape of the kernel obtained when fixing one of the radii and varying
the other is represented in Fig.\ \ref{fig:predicted}.  The minimal collision
rate is obtained for equal-size particles; this is consistent with can
be understood from the symmetry of the kernel under particle exchange.
The growth of the kernel when $a_2\gg a_1$ is essentially due to the
increase of the geometrical cross-section. Note also the presence of a
maximum when $a_2<a_1$,  attained at the crossover between
(\ref{eq:rateinterrange}) and (\ref{eq:raterange1}).  Note that a similar
shape has been proposed for the kernel in Ref.~\onlinecite{KK96}.

\begin{figure}[t!]
  \iffigs \includegraphics[scale=.4]{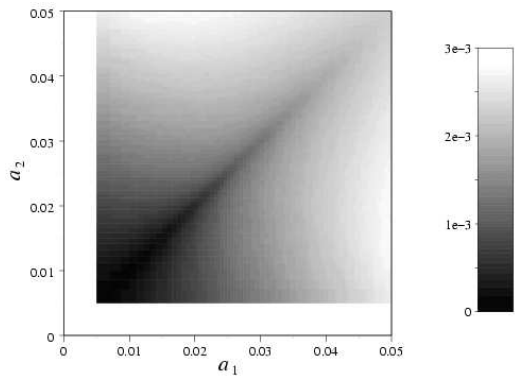}
  \else \drawing 60 10 {Fig7} \fi
  \caption{\label{fig:kernelOU} Effective inter-particle collision
    rate $Q(a_1, a_2)$ obtained numerically by considering ghost collisions in
    the case of the time-correlated random flow. The Stokes number $\St$ and
    the particle radius $a$ are related by $\St = 2 \rho_p a^2/(9 \rho_f L^2)$
    with the choice $\rho_p/\rho_f = 4.5 \times 10^{3}$. To obtain these
    rates, inter-particle statistics are computed for 100 different values of
    the Stokes number.}
\end{figure}

\begin{figure}[t!]
  \iffigs \includegraphics[scale=.4]{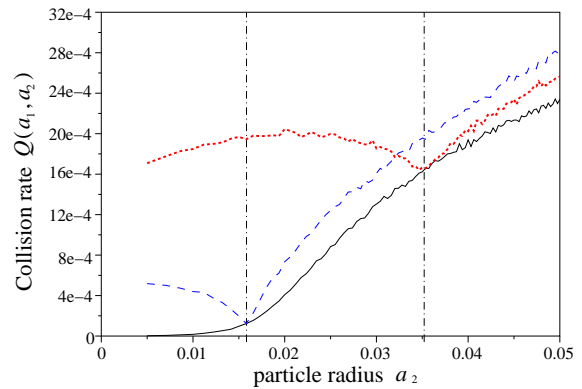} \else
  \drawing 60 10 {Fig8} \fi
  \caption{\label{fig:kernelsamples} Collision rate $Q(a_1, a_2)$
    obtained with the same settings as for Fig.\ \ref{fig:kernelOU} and
    represented here for equal-size particles, i.e. $a_1=a_2$, (solid curve,
    black online) and as a function of $a_2$ for two fixed values of $a_1$:
    $a_1=1.58\times10^{-2}$ (dashed curve, blue online), (dotted curve, red
    online) $a_1=3.54\times10^{-2}$. }
\end{figure}

The numerical results for the collision kernel $Q$ in the $(a_1,a_2)$
plane are illustrated in Fig.\ \ref{fig:kernelOU} for the Gaussian
random flow.  The one-dimensional cuts represented in
Fig.~\ref{fig:kernelsamples} compare favorably with the prediction
shown in Fig.~\ref{fig:predicted}.  The random shear flow
(\ref{eq:shear}) displays similar features (not shown).


\begin{figure}[t!]
  \iffigs \includegraphics[scale=.4]{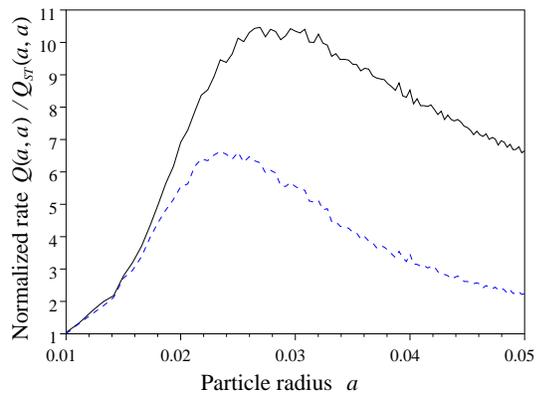} \else
  \drawing 60 10 {Fig9} \fi
  \caption{\label{fig:ratiosamestokes} Solid curve (black online): collision
    rate $Q(a, a)$ for equal-size particles, normalized by that obtained when
    neglecting particle inertia, i.e. the Saffman and Turner result
    Eq.~(\ref{eq:saffmanturner}). Dashed curve (blue online): collision rate
    obtained when neglecting correlations between the velocity difference and
    the density (i.e.\ assuming that the first is just proportional to the
    separation between the particles), normalized in the same way. These
    curves were obtained numerically with the same setting as for Fig.\ 
    \ref{fig:kernelOU}; the Saffman--Turner kernel was evaluated by following
    tracer particles advected by the same flow.}
\end{figure}

In order to quantify the importance of particle inertia the ratio
$Q(a,a)/Q_{ST}(a,a)$ between the measured kernel and that obtained
from of the Saffman--Turner approach~(\ref{eq:saffmanturner}) is
represented in Fig.\ \ref{fig:ratiosamestokes}.  To disentangle the
effects of clustering and densities-velocities correlations, the ratio
between the kernel obtained when the velocity difference is assumed to
behave linearly with the separation between the two particles, and the
Saffman--Turner prediction is represented.  Note that the two curves
coincide at very small radii: in this regime, the enhancement of
collision rates is mainly due to clustering effects.  Discrepancies
between the two curves appear rather soon, before reaching the maximum
of clustering. It is clear from Fig.\ \ref{fig:ratiosamestokes} that
in both cases, there is an increase of roughly one order of magnitude
in the collision rate of inertial particles compared with that of
tracers.  However, the measured values of the kernel differ markedly
from those obtained when only clustering effects are
considered. Therefore, away from the two asymptotics it is crucial not
to consider as independent the effects of clustering and enhanced
relative velocity.


\section{Concluding remarks}
\label{sec:conclusions}
We have identified the main mechanisms leading to the enhancement of
clustering and collisions induced by inertia in dilute suspensions of heavy
particles.  A Lagrangian method based on ghost collisions has been used for
numerical investigations.  In agreement with previous studies we found that
clustering is maximal when the Stokes number is order unity.  However, this is
not the only mechanism enhancing collisions: large velocity differences can
occur at very small separations and for finite values of the Stokes number.
This results in non-trivial scaling properties of the rate at which particles
come close.  A phenomenological model for the collision kernel based on these
ingredients has been proposed.  Our results highlight the importance of
accounting for the full position-velocity phase-space dynamics, particularly
for polydisperse solutions.  It is important to stress that, away from the
small Stokes asymptotics, multiphasic approaches,\cite{EAA83,CSTC97} based on
a continuum description of suspensions in position space, may fail to catch
these effects.

We conclude by discussing some aspects concerning impurities in
turbulent flows.

We expect our results to be relevant to clustering and collisions at
dissipative scales. Turbulence will certainly affect both the values
of the constants and the scaling exponents, and induce a non-trivial
Reynolds number dependence of the collision rates.  Within the
framework of model flows it would be interesting to extend the present
study to random flows with non-Gaussian statistics to account for
intermittency of the velocity gradients. Moreover, in actual turbulent
flows the Kolmogorov time-scale is a dynamical variable which has
important fluctuations.  Therefore, the Stokes number is also a random
variable along the particle trajectories.  This provides a further
complication in developing models accounting for clustering and
collisions of particle suspensions, which needs to be investigated.

Another important issue concerns clustering at inertial-range scales,
where the velocity field is not differentiable.  Experiments indeed
show that preferential concentration appears also at those
scales.\cite{EF94} Non-trivial clustering properties have been
observed numerically also in the inverse-cascade range of
two-dimensional turbulent flows, namely the formation of holes in the
distribution of particles.\cite{BLG04} It is important to remark that
clustering at the inertial scales may influence the probability for
two particles to arrive below the dissipative scale and thus the
collision rates.  In the inertial range the dynamics of the fluid is
close to Kolmogorov 1941 theory, i.e.\ the velocity field is H\"older
continuous with exponent $1/3$. As a consequence, tracers separate
explosively giving rise to the celebrated Richardson's $t^{3/2}$
law. For inertial particles, one needs to understand the competition
between explosive separation and clustering due to dissipative
dynamics.  In this direction, it may be useful to further extend to
inertial particles recent models and techniques developed in the
framework of passive scalars (for two recent reviews see
 Refs.~\onlinecite{SS00,FGV01}).

\section*{Acknowledgments}

We thank A. Vulpiani for motivating us to start this study.  We warmly
acknowledge F.~Cecconi and B.~Marani for their continuous
support. A.C. acknowledges hospitality and support from the INFM
Center SMC. This work was supported by the European Union Network
``Fluid mechanical stirring and mixing'' under contract
HPRN-CT-2002-00300.


\begin{thebibliography}{99}
\bibitem{nota}This requires to assume
ergodicity for the particle dynamics, which is usually the case.
\end{thebibliography}
\end{document}